# Optimal Planning for Electrical Collector System of Offshore Wind Farm with Double-sided Ring Topology

Xinwei Shen, *Senior Member*, *IEEE*, Qiuwei Wu, *Senior Member*, *IEEE*, Hongcai Zhang, *Member, IEEE,* Liming Wang, *Senior Member, IEEE*

*Abstract*—We propose a planning method for offshore wind farm electrical collector system (OWF-ECS) with double-sided ring topology meeting the "N-1" criterion on cable faults, in which the submarine cables layout of OWF is optimized considering cable length and power losses. The proposed mixed-integer quadratic programming (MIQP) model is based on the Capacitated Vehicle Routing Problem (CVRP) formulation and power network expansion planning, which could approximate the power losses in OWF-ECS. In addition, cross-avoidance constraints are proposed to avoid crossing cables, and the minimum k-degree center tree model is included to improve the convergence. Case studies on OWFs with 30 and 62 WTs demonstrate the effectiveness of the proposed method. Considering the potential outage cost in the radial topology, the total cost of the planning result is reduced by up to 25.9% with reliability improvement. The cable investment is reduced by 4%~8% with the proposed method compared with conventional heuristic methods and Google OR-tools. The proposed method/model can also achieve acceptable computation efficiency and OWF-ECS planning results with good optimality. Moreover, it could be solved by modern commercial solvers/optimization software, thus it's easy to use even for large-scale OWF.

*Index Terms*—Electrical collector system, mixed integer programming, N-1 criterion, offshore wind farm

## NOMENCLATURE

**Sets and parameters**:

| | |
|---|---|
| $V$ | node set, with $|V|$ nodes |
| $V^{Sub}/V^W$ | substation/wind turbine node set |
| $L$ | cable set, with $|L|$ cables |
| $L^j$ | cables connected with node $j$ |
| $L^{Sub}$ | cables connected with substation node |
| $(i,j)$ | a cable between nodes $i$ and $j$ |
| $r_{ij}$ | resistance of cable $(i,j)$ |
| $P_j^W$ | wind power generation at node $j$ |
| $c_{ij}$ | cost of cable $(i,j)$, unit: ￥ |
| $\bar{P}_{ij}$ | the transmission capacity of cable $(i,j)$ |
| $\eta$ | coefficient of power loss to the whole life-time (planning years) |
| M | big-M constant |
| $B_{ij}$ | susceptance of cable $(i,j)$ |
| $S$ | Set of cables in the minimum k-degree centre tree (k-DCT) |
| $S^0/S^1$ | Set of cables not in the k-DCT adjacent/not adjacent to the substation node |
| $R_{\min}$ | Minimum cable routes needed to transmit all power of WTs |
| $Y^0$ | Number of cables adjacent to substation node but not in the k-DCT |

**Variables**:

| | |
|---|---|
| $x_{ij}$ | investment decision variable for adding cable $(i,j)$, $x_{ij}=1$ if $(i,j)$ is selected, else $x_{ij}=0$ |
| $x_{ij}^+/x_{ij}^-$ | decision variable denoting the direction of power flow in CVRP model. $x_{ij}^+=1$ ($x_{ij}^-=1$) if the power flow moves from node $i$ ($j$) to node $j$ ($i$) in cable $(i,j)$ |
| $u_j$ | the accumulated power collected by submarine cables when arriving at node $j$ in CVRP model |
| $P_{ij}$ | power flow/current in $(i,j)$ |
| $p_j^{shed}$ | curtailed wind power at node $j$ |
| $P_j^{Sub}$ | power collected in offshore substations |
| $\theta_j$ | phase angle of node $j$ |
| $y_{ij}$ | decision variable denoting if cable $(i,j)$ is in the k-DCT, $y_{ij}=1$ if $(i,j) \in S$, otherwise $y_{ij}=0$ |
| $y_{ij}^0/y_{ij}^1$ | decision variable denoting if cable $(i,j)$ not in the k-DCT is in $S^0(S^1)$, $y_{ij}^0=1$ ($y_{ij}^1=1$) if $(i,j) \in S^0(S^1)$, otherwise $y_{ij}^0=0$ ($y_{ij}^1=0$) |

## I. INTRODUCTION

O cean renewable energy has attracted more and more attention these years with rising concerns on the carbon emissions and climate change globally. Offshore wind farms (OWFs) are the most promising and commercially successful one compared with other ocean renewable technologies, e. g., wave energy, tidal energy, etc. [1]. According to the Global Wind Energy Council's report in 2021[2], the sharp drop of levelized cost of energy of the OWF makes it one of the most competitive energy sources, and it increasingly plays a unique role in facilitating cross industry cooperation and decarbonization. By the end of 2020, the total installed capacity of the OWF was around 35 GW. It will be about 270 GW by 2030 and 2,000 GW by 2050 to achieve net zero emission. For most of the OWFs, the utilization time of more than 3000 hours per year is expected [3], significantly higher than that of onshore wind farms. Therefore, it compensates for the additional costs of offshore plants to some extent. In recent years, more and more large-scale OWFs have been constructed,

This work was supported in part by National Natural Science Foundation of China (No. 52007123). Dr. Xinwei Shen, Dr. Qiuwei Wu and Dr. Liming Wang are with Tsinghua Shenzhen International Graduate School, Tsinghua University, Shenzhen, 518055, China. Dr. Hongcai Zhang is with the State Key Laboratory of Internet of Things for Smart City, University of Macau. (*Corresponding*: Xinwei Shen, sxw.tbsi@sz.tsinghua.edu.cn)

e.g., in France, the Saint-Brieuc OWF[5] is a large-scale OWF with 496 MW total installed capacity, in which 62 8 MW wind turbines (WTs) are installed, and 90 km 66 kV cables are utilized to collect all the electric energy from WTs. However, various operation conditions and extreme weather in the offshore areas pose a higher economic and reliability requirement on OWFs as the consequence of faults is more significant [4].

The OWF's electrical collector system (ECS) planning and design are very important because changing cabling topology can improve the system reliability, e.g., single/double-sided ring topology, as shown in **Fig. 1** [6], while the investment and power loss in the ECS can also be reduced by optimizing the cabling design. In this regard, many scholars have proposed planning methods for the OWF ECS.

In [7], the minimum spanning tree (MST) algorithm was used in the radial ECS planning for wind farm, in which the total trenching length is optimized and further reduced by introducing intermediate splice points similar to the Steiner's vertices. However, the Steiner points are usually not allowed in the OWF ECS because branching a marine cable would be very vulnerable. Reference [8] also focuses on radial ECS planning for the OWF, in which the power losses are further incorporated by a precomputing strategy. Thus, the proposed model is still a Mixed-Integer Linear Program (MILP). The authors also discuss how to include and improve the no-cross constraints as it's the restriction for the OWF ECS. Similarly, [9] proposed an MILP model for the OWF ECS planning with a radial structure, in which the total power losses are also included in the objective function with a pre-processing strategy. But the power flow model and pre-processing method are not applicable for the OWF ECS with a ring topology. A two-stage approach is proposed in [10] for the onshore wind farm ECS with a radial topology. The proposed method minimizes the total trenching length between wind turbines and substation in the first stage and determines the cabling in the second stage. It is shown that the optimal cable selection for any line can be known in advance. In [11], a bi-level multi-objective optimization framework is proposed for the OWF ECS planning, the configuration of WTs, i.e., the installation capacity and positioning of WTs, is simultaneously designed with the ECS topology for the OWF. However, methods in [10] and [11] are not applicable for the ring topology.

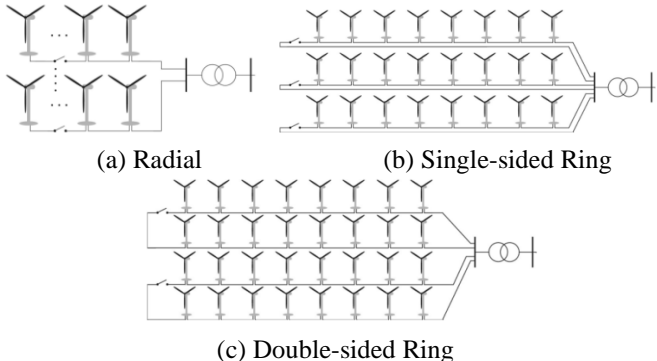

(a) Radial            (b) Single-sided Ring

(c) Double-sided Ring

**Fig. 1**. Topology of OWF ECS: (a) Radial (b)Single-sided Ring (c) Double-sided Ring

Considering the failure rate of facilities in OWF and the long mean time to repair (MTTR), it's essential to study the ring (loop) structure in the OWF ECS planning, because any cable outage can result in lots of wind energy losses with the radial design. In fact, it's reported in [19] that the failure rate of 1 km MV marine cable in the OWF is 0.015 per year and the MTTR could be as high as 1440 hours (60 days). In [12], three cabling structures for the OWF are explored: the string (radial), ring, and multiloop structure. it's demonstrated that the multiloop structure increases reliability and proves to be the most economic when the failure rate and MTTR of cables are relatively high. However, the produced multiloop structure is based on choosing "connecting points" in the ring structure. In [13], the authors used the "cross-substation incorporation" to the OWF ECS, i.e., capacity sharing among substations, to reduce the capacity requirement for transformers in each substation, but related reliability enhancement benefits are not discussed. Consequently, it cannot produce the optimal multiloop structure for the ECS of the wind farm. In [14] and [18], the double-sided ring topology was used in the OWF, but the designed ECS obviously doesn't meet the "N-1" criterion, and there is no guarantee on the solution optimality for the planning result obtained by the CWS and GA algorithms.

As for modeling and solution methods, the ECS planning for both radial and double-sided ring topology was often modeled as the well-known capacitated vehicle routing problem (CVRP) [12][13][14]. Some researchers also propose to combine micro-siting and ECS planning in OWF[11][15]-[17]. However, most of the solving algorithms are based on the heuristic algorithms, e.g., the MST [7][10][13], Sweep and Clarke and Wright savings (CWS) [12][13][14], NSGA-III[11], PSO[10][11][15], genetic algorithm[16], hybrid grey wolf optimization (HGWO)[17] etc. The convergence performance with the heuristic algorithms is case-by-case, and the optimality gap between their final solutions and the global optimal solution cannot be obtained [20].

Moreover, the power network expansion planning (PNEP) problem, including transmission/distribution network, has been studied for decades, in which the mathematical optimization models considering the topology of the network[23],[24] and power supply reliability[25] are discussed, explored and applied comprehensively, showing its great benefits in reducing power system power losses, saving investment costs on power lines/cables, as well as improving reliability.

The existing research focuses on the design of the radial ECS for OWF, while some of them rely on the heuristic algorithms to obtain the ring topology design. Moreover, none of them considers the "N-1" criterion for cable outages.

In this paper, we propose a planning method for the OWF-ECS planning (OWF-ECSP) problem with the double-sided ring topology, based on the CVRP and ideas from the PNEP. The contributions are summarized as below:

1)    An optimal planning scheme for the OWF ECSP with the double-sided ring topology is proposed, in which a mixed-integer quadratic programming (MIQP) model is developed to optimize the cabling based on the CVRP model and the PNEP model. It considers "N-1" security criterion in submarine cable connections, minimization of trenching length and power losses throughout the lifetime of the OWF.

2) To meet the requirement of the OWF-ECSP, crossing-avoidance-constraints (CAC) are proposed to avoid the crossing cables in OWF ECS. Numerical results show that with CAC the proposed method outperforms some other solvers solving CVRP (e.g., Google OR-tools) as it could produce rational ECSP result while maintaining acceptable computation efficiency.

3) To improve the computation deficiency, we propose to include the k-degree center tree (k-DCT) as a complementary part of the model. Numerical results show that it helps the model obtain a better lower bound and thus improving the optimality gap of final results.

The remaining parts of the paper is structured as follows: Section II describe the procedure of proposed OWF-ECSP method, Section III formulates the mathematical model, case studies are given in Section IV, Section V concludes the paper.

## II. PROPOSED OWF-ECSP METHOD

### A. Framework of Proposed Method

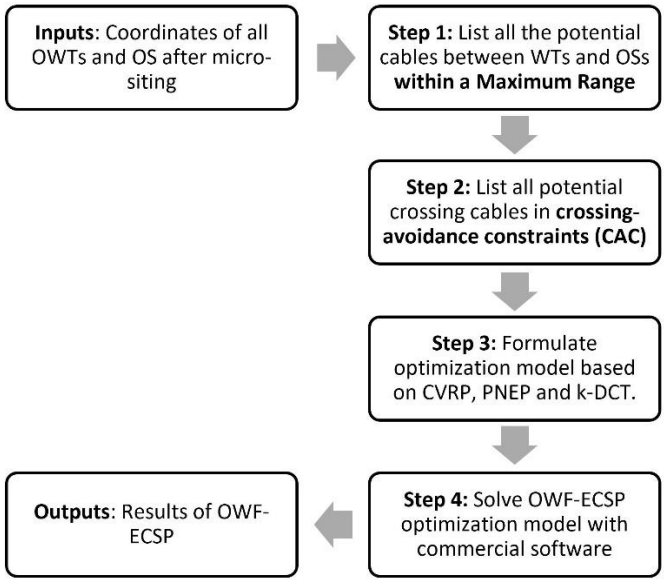

**Fig. 2**. Proposed OWF-ECSP methods' framework

The proposed OWF-ECSP method is divided into several steps. The framework of the proposed method is illustrated in **Fig. 2**.

Firstly, the inputs are the coordinates of all WTs and offshore substation (OS) determined by the micro-siting for the OWF [11][22], e.g., using the fuzzy c-means method to locate the OS [13][14]. Since the focus of the proposed method is to improve the optimality of OWF ECS planning result, especially the cabling topology, it's too complicated to further co-optimize the layout of WTs considering the Wake Effect in micro-siting even though it's also important for the OWF. The voltage level, as well as the number/location/connection form of offshore substations etc., is assumed to be pre-determined too, even though they also influence the total cost of ECS.

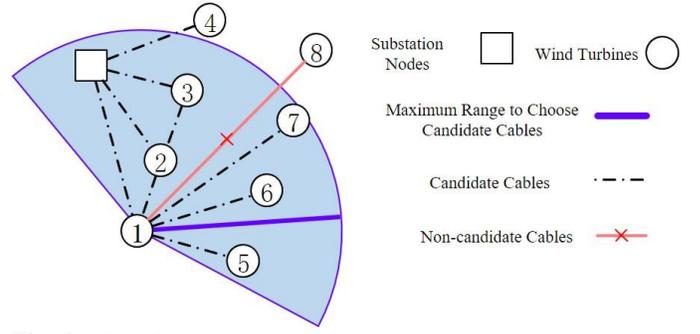

**Fig. 3.** Step 1: list all the potential connection cables between neighborhood WTs and OSs

In **Step 1**, list all the potential connection cables between neighboring WTs and OSs *within a pre-determined maximum range*, as illustrated in Fig. 3. As such, the optimal cabling layout doesn't require long cables between WTs and OSs, e. g., cables crossing more than 3 WTs. Similar ideas can be found in the *Granular Tabu Search* for the CVRP [34].

In Fig. 3, the potential cables to connect WT 1 is 1-2, 1-5, 1-6, 1-7 and 1-Sub, while 1-4 and 1-8 are not candidate cables because they are longer than the *maximum range*. The same rules also apply to other WTs to list all the candidate cables. It's worth noting that, candidate cables for substations can be specially designed instead of only dependent on the distance. An example of the candidate cables for an OWF with 30 WTs is illustrated in Fig. 4. The blue square node is the OS node, the yellow round nodes are WT nodes, and the dotted lines are candidate cables.

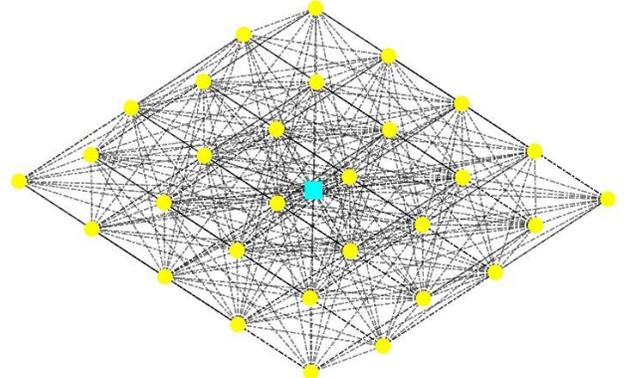

**Fig. 4**. Candidate cables for a 30-wind-turbines (30-WTs) OWF

In **Step 2**, list all crossing candidate cables, which can be put into the constraints to avoid cables crossings. The details of this step are discussed in the next section. Sometimes, this step can be omitted because line crossings do not appear in the planning results.

In **Step 3**, formulate the model based on CVRP, PNEP and k-DCT, which are all MIP models as described in Section III. The model only obtains the planning result with the double-sided ring topology, which meets the "N-1" criterion naturally.

In **Step 4**, the model is solved by the off-the-shelf commercial software, e. g. CPLEX[29] or GUROBI[30], which is able to handle large-scale MIP problems, then produces the planning result.

## B. Cross-avoidance Constraints in Step 2

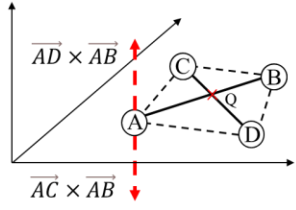

**Fig. 5**. An example of crossing cables' pair (AB, CD)

Connecting WTs in a ring structure is similar to the euclidean Traveling Salesman Problem (TSP) [8]. As Fig. 5 shows, consider node A to be the substation node (or the starting node of the salesman), B/C/D are WTs (or cities to be traveled by the salesman), they are almost at the same height above the sea level, constituting the plane of ABCD. The candidate traveling salesman's routes, e.g. {AD, AB, CD, BD}, are like cables connecting WTs and substations, which usually lie on the seabed of the OWF but could be illustrated by the line segments on the plane ABCD.

To find out the crossing cables' pairs and include them in **crossing-avoidance constraints (CAC)**, we develop a method to judge whether each pair of cables is crossing each other, e.g., (AB, CD) as shown in **Fig. 5**. It can be seen from **Fig. 5** that, nodes C and D are on both sides of line AB, while nodes A and B are on both sides of line CD. As a result, according to the "right-hand rule" of vector cross product[26], $\overrightarrow{AC} \times \overrightarrow{AB}$ ($\overrightarrow{CB} \times \overrightarrow{CD}$) is perpendicular to ABCD plane and in the negative direction, while $\overrightarrow{AD} \times \overrightarrow{AB}$ ($\overrightarrow{CA} \times \overrightarrow{CD}$) is in the positive direction. Therefore, line segments AB and CD intersect if and only if

$$\left(\overrightarrow{AC} \times \overrightarrow{AB}\right) \cdot \left(\overrightarrow{AD} \times \overrightarrow{AB}\right) < 0 \text{ and}$$
$$\left(\overrightarrow{CB} \times \overrightarrow{CD}\right) \cdot \left(\overrightarrow{CA} \times \overrightarrow{CD}\right) < 0$$

This is the condition to find out the crossing cables' set denoted by $L^{CAC}$. This condition is used to traverse all pairs of candidate cables to constitute $L^{CAC}$, namely,

$$L^{CAC} = \{(AB, CD) | \left(\overrightarrow{AC} \times \overrightarrow{AB}\right) \cdot \left(\overrightarrow{AD} \times \overrightarrow{AB}\right) < 0$$
$$\text{and } \left(\overrightarrow{CB} \times \overrightarrow{CD}\right) \cdot \left(\overrightarrow{CA} \times \overrightarrow{CD}\right) < 0, \forall A, B, C, D \in V\}$$

Then assuming $x_{ij}$ is the binary variable denoting whether line $(i, j)$ between node $i$ and $j$ should be invested, we have,

$$x_{AB} + x_{CD} \leq 1$$

which means cables AB and CD can't be invested at the same time in the final planning result. This constraint is applied to all cable pairs in $L^{CAC}$.

We wish to emphasize that, it's not necessary to list all the crossing cables' pairs in $L^{CAC}$, because in the optimal planning result for a TSP, it's likely to avoid crossing route due to triangle's feature (in the triangle AQD and BQC, the sum of any two sides has Euclidean length greater than the third side, thus A-D-B-C-A is better than A-B-D-C-A on the left). However, we propose to include cables connecting substation with other WTs to avoid crossing. Because if the ring topology is too long, there's still possibility of crossing cables.

## III. OWF-ECSP PROBLEM FORMULATION

### A. CVRP-based Model with Double-sided Ring Topology

As mentioned earlier, the well-known CVRP model can be modified for the OWF-ECSP. Assume a complete weighted directed graph $G = (V, L)$, where $V = \{V^{Sub}, V^W\}$ includes substation nodes in $V^{Sub}$ and WT nodes in $V^W$, $L$ denotes all the candidate cables connecting the nodes, $L = \{(i, j) \in V \times V : i \neq j\}$, and each candidate cable $(i, j)$ is associated with a cost $c_{ij}$ and a binary investment decision variable $x_{ij}$. Here the vehicle-flow formulation [27] is used for the directed CVRP to be a model for the OWF-ECSP,

$$\min_{x,P} \sum_{(i,j) \in L} c_{ij} x_{ij} + \eta \sum_{(i,j) \in L} r_{ij} P_{ij}^2 + M \sum_{i \in V^W} P_i^{shed} \quad (1)$$

where the objective function (1) aims to minimize the cost of constructing ECS cables, power losses and curtailed wind power. This is a multi-objective problem, in which the summation of each term's equivalent value is presented and thus converted to a single objective. The cable cost is denoted by summing the cost of each cable $c_{ij}$ multiplied with its investment decision $x_{ij}$, i.e. $\sum_{(i,j) \in L} c_{ij} x_{ij}$. The power losses cost throughout the lifetime of the OWF is denoted by $\eta \sum_{(i,j) \in L} r_{ij} P_{ij}^2$, where $\eta$ is the coefficient to transfer power loss at one time slot to the whole lifetime of the OWF, i.e., $\eta$=planning years (e.g., 20 years)×yearly full load hours (e.g., 3000 hours). It should be noted that, since $|U_i| \approx 1$ p. u. in the OWF's ECS, $I_{ij} = P_{ij}/U_i \approx P_{ij}$, thus we approximate the power loss $P_{ij}^{loss} = r_{ij} I_{ij}^2 \approx r_{ij} P_{ij}^2$. Moreover, the total value of the curtailed wind power $M \sum_{i \in V^W} P_j^{shed}$ is also included, but usually not allowed to be nonzero since M is a big constant, e.g., $10^7$¥/MWh.

The CVRP-related constraints are as follows.

$$x_{ij}^+ + x_{ij}^- = x_{ij} \quad \forall (i, j) \in L \quad (2)$$

$$\sum_{(i,j) \in L} x_{ij}^+ + \sum_{(h,i) \in L} x_{hi}^- = 1 \quad \forall i \in V^W \quad (3)$$

$$\sum_{(i,j) \in L} x_{ij}^- + \sum_{(h,i) \in L} x_{hi}^+ = 1 \quad \forall i \in V^W \quad (4)$$

$$u_i - u_j + P^W \leq \bar{P}_{ij}(1 - x_{ij}^+) \quad \forall (i, j) \in L \quad (5)$$

$$u_j - u_i + P^W \leq \bar{P}_{ij}(1 - x_{ij}^-) \quad \forall (i, j) \in L \quad (6)$$

$$P^W \leq u_i \leq \bar{P}_{ij} \quad \forall i \in V^W \quad (7)$$

In (2), both $x_{ij}^+$ and $x_{ij}^-$ are binary variables, $x_{ij}^+ = 1$ ($x_{ij}^- = 1$) if the fictious power flow moves from node $i$ ($j$) to node $j$ ($i$) in cable $(i, j)$, which could only happen if it's invested ($x_{ij} = 1$). Eqs. (3) and (4) denote that the outflow and inflow cable of each WT node is exactly 1, which is a requirement for the double-sided ring topology in the OWF, and also similar to the CVRP rule. Eqs. (5) and (6) are known as the MTZ-formulation or MTZ-specific Subtour Elimination Constraints [28]. The variables $u = \left(u_1, \dots, u_{|V^W|}\right)^T$ indicate the accumulated power $u_i$ already collected by submarine cables when arriving at node $i$, thus it should be limited by line capacity $\bar{P}_{ij}$ as shown in (7). A simple example is shown in **Fig. 6**.

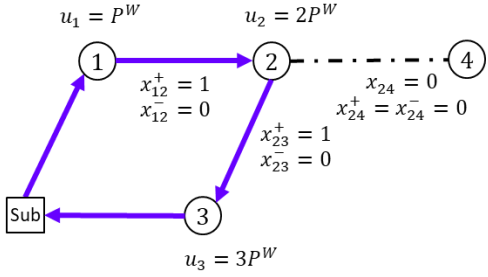

**Fig. 6**. A simple example of CVRP-based model

In this example, the route "Sub→1→2→3→Sub" is a double-sided ring connecting substation node with WT nodes 1, 2, and 3. As can be seen, the accumulated power $u_i$ increase along with the route from $u_1 = P^W$ to $u_3 = 3P^W$, but it's possible only if $x_{12}^+ = 1$ and $x_{23}^+ = 1$. The cable (2,4) is not invested because Eqs. (3) and (4) restrain that the outflow/inflow cable of node 2 is exactly 1, thus $x_{24} = 0$ and WT node 4 is left for another route to connect. If the OWF has more than 1 offshore substation, planning model based on multi-depot CVRP[31] should be developed, but it would change the model formulation significantly, thus it's out of the scope of the paper and considered as a future research topic.

As mentioned earlier, the crossing-avoidance constraints eliminating all the possible crossing cables combinations $(i_1 j_1, i_2 j_2) \in L^{CAC}$ in the planning results are also included:

$$x_{i_1 j_1} + x_{i_2 j_2} \leq 1 \quad \forall (i_1 j_1, i_2 j_2) \in L^{CAC} \tag{8}$$

In fact, the planning results obtained by the CVRP-based model meets the "N-1" criterion naturally. Because if any single cable outage happens in any "vehicle route", the wind power could still be transmitted to substation by the other way around, since in each "vehicle route" the accumulated power $u_i \leq \bar{P}_{ij}$.

### B. DC-power-flow-based Transmission Planning Model

Beside the CVRP model, the power flow in each submarine cable is needed to approximate and optimize the total power losses in the OWF ECS. Therefore, a MILP-based transmission planning model is adopted, in which the optimal power flow is based on DC power flow, as follows.

$$\sum_{(i,j) \in L^j} P_{ij} = P_j^W - P_j^{shed} \quad \forall j \in V^W \tag{9}$$

$$\sum_{(i,j) \in L^j} P_{ij} = P_j^{Sub} \quad \forall j \in V^{Sub} \tag{10}$$

$$\left| B_{ij}(\theta_j - \theta_i) - P_{ij} \right| \leq (1 - x_{ij})M \quad \forall (i,j) \in L \tag{11}$$

$$\theta_i = 0 \quad \forall j \in V^{Sub} \tag{12}$$

$$-x_{ij}\bar{P}_{ij} \leq P_{ij} \leq x_{ij}\bar{P}_{ij} \quad \forall (i,j) \in L \tag{13}$$

$$P_i^W \geq P_i^{shed} \geq 0 \quad \forall i \in V^W \tag{14}$$

where Eq. (9) is the power balance at WT node $j \in V^W$, $L^j$ denotes the cables set connecting node $j$, thus $\sum_{(i,j) \in L^j} P_{ij}$ is the total transmitted power by cables connecting node $j$, which should be equal to the wind generation $P_j^W$ minus curtailed wind $P_j^{shed}$. Similarly, Eq. (10) is the power balance at substation node $j \in V^{Sub}$, with $P_j^{Sub}$ denoting the power collected from all the cables. Eq. (11) denotes the relation of the phase angle $\theta_i$ and $\theta_j$ in cable $(i,j)$ once the cable is invested ($x_{ij} = 1$) with big-M method, while Eq. (12) shows the slack bus (reference node of phase angle) is the substation node. Eqs. (13) and (14) limit the transmitted power $P_{ij}$ and curtailed wind power $P_i^{shed}$, respectively.

### C. Minimum k-Degree Centre Tree Model

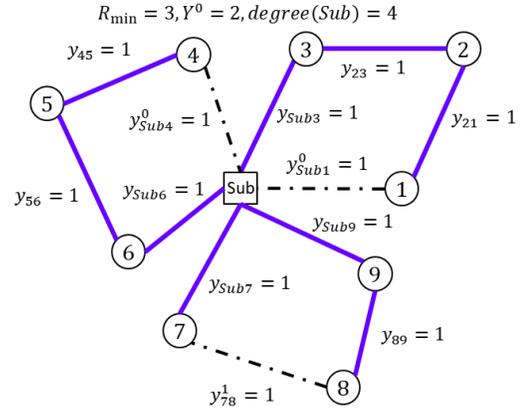

**Fig. 7**. A simple example of a k-DCT

To improve the convergence of large-scale OWF ECSP, we develop and incorporate a model known as the minimum k-degree centre tree (k-DCT)[32]. Consider a OWF ECSP result with several cable routes, it can also be treated as the solution to a multiple TSP (M-TSP), which is apparently a lower bound of CVRP since the CVRP is the M-TSP with additional constraints on cable capacity. The minimum number of cable routes needed to transmit all wind power $R_{min}$ is calculated by

$$R_{min} = \left\lceil \frac{|V^W|P^W}{\bar{P}_{ij}} \right\rceil \tag{15}$$

where the $\lceil \cdot \rceil$ means rounded up. If we "remove" (not actually remove, just assume for analysis) $Y^0 \leq R_{min}$ cables from $L^{Sub}$ and $R_{min} - Y^0$ cables from $L - L^{Sub}$, then $R_{min}$ cables are removed from the planning result, i.e. one from each cable routes, it produces a tree where the degree of substation node is $k = 2R_{min} - Y^0$, namely a k-DCT.

A simple example with $R_{min} = 3$, $Y^0 = 2$, $degree(Sub) = 4$ is shown in **Fig. 7**. There are $R_{min} = 3$ cable routes in it, assume that we "remove" cable $(sub, 4)$ and $(sub, 1)$, $Y^0 = 2$ cables from $L^{Sub}$, and "remove" cable $(7,8)$ from $L - L^{Sub}$, a k-DCT is obtained, where the degree of the substation node is 4.

The k-DCT is formulated as follows.

$$x_{ij} = y_{ij} + y_{ij}^0 + y_{ij}^1 \quad \forall (i,j) \in L \tag{16}$$

$$\sum_{(i,j) \in L^{Sub}} y_{ij} = 2R_{min} - Y^0 \tag{17}$$

$$\sum_{(i,j) \in L} y_{ij} = |V^W| \tag{18}$$

$$\sum_{(i,j) \in L^{Sub}} y_{ij}^0 = Y^0 \tag{19}$$

$$\sum_{(i,j)\in L-L^{sub}} y_{ij}^1 = R_{\min} - Y^0 \quad (20)$$

$$\sum_{(i,j)\in Lj} x_{ij} = 2 \quad \forall j \in V^W \quad (21)$$

Eq. (16) denotes the relationship between cable investment decisions and the k-DCT, as well as the "removed" cables, Eq. (17) imposes the degree on substation node, Eqs. (18)[19] and (20) specify the number of cables required, Eq. (21) imposes the degree of each WT node in M-TSP solution is 2.

Note that introducing $y_{ij}, y_{ij}^0, y_{ij}^1$ and k-DCT doesn't affect the feasibility of any solution to OWF ECSP. However, it can provide a tight lower bound for the planning problem, thus helping the convergence[32].

Therefore, the compact form of the OWF ECSP model is,

$$\min_{x,P} (1)$$

$$\text{s. t. (2)- (21)}$$

which is an MIQP that can be solved by modern branch-and-cut solvers[29][30].

### D. Reliability Assessment for radial OWF ECS

In addition to minimizing the total cable length and power losses in (1), the radial OWF's reliability is also evaluated by calculating the cost of expected energy not generated (EENG) considering the outage probability and MTTR of submarine cables, defined as,

$$C^{EENG} = Pr^{ele} \cdot MTTR \cdot py \cdot fr \cdot \sum_{(i,j)\in L} l_{ij} x_{ij} P_{ij} \quad (22)$$

where $C^{EENG}$ denotes the cost of EENG, $Pr^{ele}$ is the price of wind energy (unit: ¥/MWh), MTTR is the mean time to repair (unit: hour), $py$ is the lifetime of OWF (unit: year), e.g., 20 years, $fr$ denotes the failure rate of submarine cables in the OWF (unit: time/km per year), and $l_{ij}$ denotes the length of cable $(i,j) \in L$. Thus, $fr \sum_{(i,j)\in L} l_{ij} x_{ij} P_{ij}$ denotes the EENG considering whether or not the submarine cable is invested ($x_{ij} = 1 \ or \ 0$) and corresponding power transmitted in each cable is $P_{ij}$.

## IV. CASE STUDIES

We design 4 cases to show the effectiveness of the proposed method:

**Case 1:** OWF-ECSP with the radial topology[21];
**Case 2:** OWF-ECSP based on the Sweep + CWS[12][13][14];
**Case 3:** OWF-ECSP solved by Google OR-tools[37];
**Case 4:** OWF-ECSP based on proposed method.

In Case 1, we only consider the radial topology in the OWF-ECSP. Therefore, it obviously doesn't meet "N-1" criterion and the aforementioned reliability assessment should be applied to evaluate the cost of the EENG. The failure rate is set as $fr = 0.0045$ time/km · yr, according to the MVAC transmission system failure rate of OWFs in [38], while the $MTTR$ is set to 1440 hours as in [12], $Pr^{ele} = 850$ ¥ /MWh, and $py = 20$ years. In Cases 2, 3 and 4, the double-sided ring topology are considered. Thus, it's not necessary to evaluate the cost of the EENG since they naturally meet the "N-1" criterion as mentioned at the end of Section III. A. The traditional heuristic method based on the Sweep+CWS, which is applied in [12]-[14], is used in Case 2. The "sweep" process starts from wind turbine No. 1 and is in the counter clockwise direction in both examples. Google OR-tools is used in Case 3 to solve corresponding CVRP. Case 2 and 3 are used to be compared with the proposed method in Case 4.

In Cases 1, 2 and 4, we solved the OWF-ECSP model by the Gurobi and MATLAB on a desktop PC with Intel Core i7-1165G7. Since there's not a standard IEEE test case for OWF-ECSP research yet, we utilized a fictitious 30-WTs OWF and the previously-mentioned 62-WTs Saint-Brieuc OWF [5], with other case conditions the same as the real-world applications, e.g., the distance between each row and column, and voltage level [35][36]. The case conditions are summarized below.

TABLE I
CASE CONDITIONS OF 30/62 WTS EXAMPLE

| Num of WTs | Capacity of each WT (MW) | Distance between each row (km) | Distance between column (km) | Maximum capacity of cable (MW) | Base voltage level (kV) | $R_{\min}$ |
|---|---|---|---|---|---|---|
| 30 | 5 | 1 | 1.3 | 32 | 35 | 5 |
| 62 | 8 | 1 | 1.3 | 65 | 66 | 8 |

$R_{\min}$ is calculated with (15), and we take $Y^0 = R_{\min}$ in both examples. The other data of case studies, including the coordinates of all WTs as inputs, as well as the cost of candidate submarine cables, can be found in [39]. The full load hours in these OWFs are assumed to be 4000 hr. The big M in (1)/(11) is set to $10^7$ because smaller M could cause wind curtailment in planning results and no acceleration could be guaranteed according to a sensitivity analysis for M. In both examples, there's only 1 offshore substation, because OWF with multiple substations is out of scope of the proposed method and would be discussed and addressed in future research.

### A. Planning Results Analysis

Table II
OWF-ECSP RESULTS: (A) 62-WTs (B) 30-WTs EXAMPLE

(a)

| Num of WTs | 62 | | | |
|---|---|---|---|---|
| | Case 1 | Case 2 | Case 3 | Case 4 |
| Inv. Cost (¥M) | 284.3 | 415.9 | 399.3 | 386.0 |
| Ope. Cost (¥M) | 51.6 | 30.0 | 30.2 | 28.8 |
| Cost of EENG | 224.1 | 0 | 0 | 0 |
| Total Cost | 560.0 | 445.8 | 429.4 | 414.8 |
| $P_{loss}$ rate (%) | 0.159 | 0.089 | 0.089 | 0.086 |
| Opt. Gap* (%) | 5.92 | None | None | 5.00 |

(b)

| Num of WTs | 30 | | | |
|---|---|---|---|---|
| | Case 1 | Case 2 | Case 3 | Case 4 |
| Inv. Cost (¥M) | 130.4 | 187.2 | 178.2 | 180.3 |
| Ope. Cost (¥M) | 16.7 | 10.5 | 10.4 | 10.1 |
| Cost of EENG | 46.2 | 0 | 0 | 0 |
| Total Cost | 193.3 | 197.7 | 188.7 | 190.4 |
| $P_{loss}$ rate (%) | 0.164 | 0.10 | 0.10 | 0.10 |
| Opt. Gap* (%) | 0.98 | None | None | 1.00 |

*Optimality Gap: the relative gap between upper bound (obj. value of current best feasible solution) and lower bound (obj. value of relaxed problem) when we terminated the solving process.

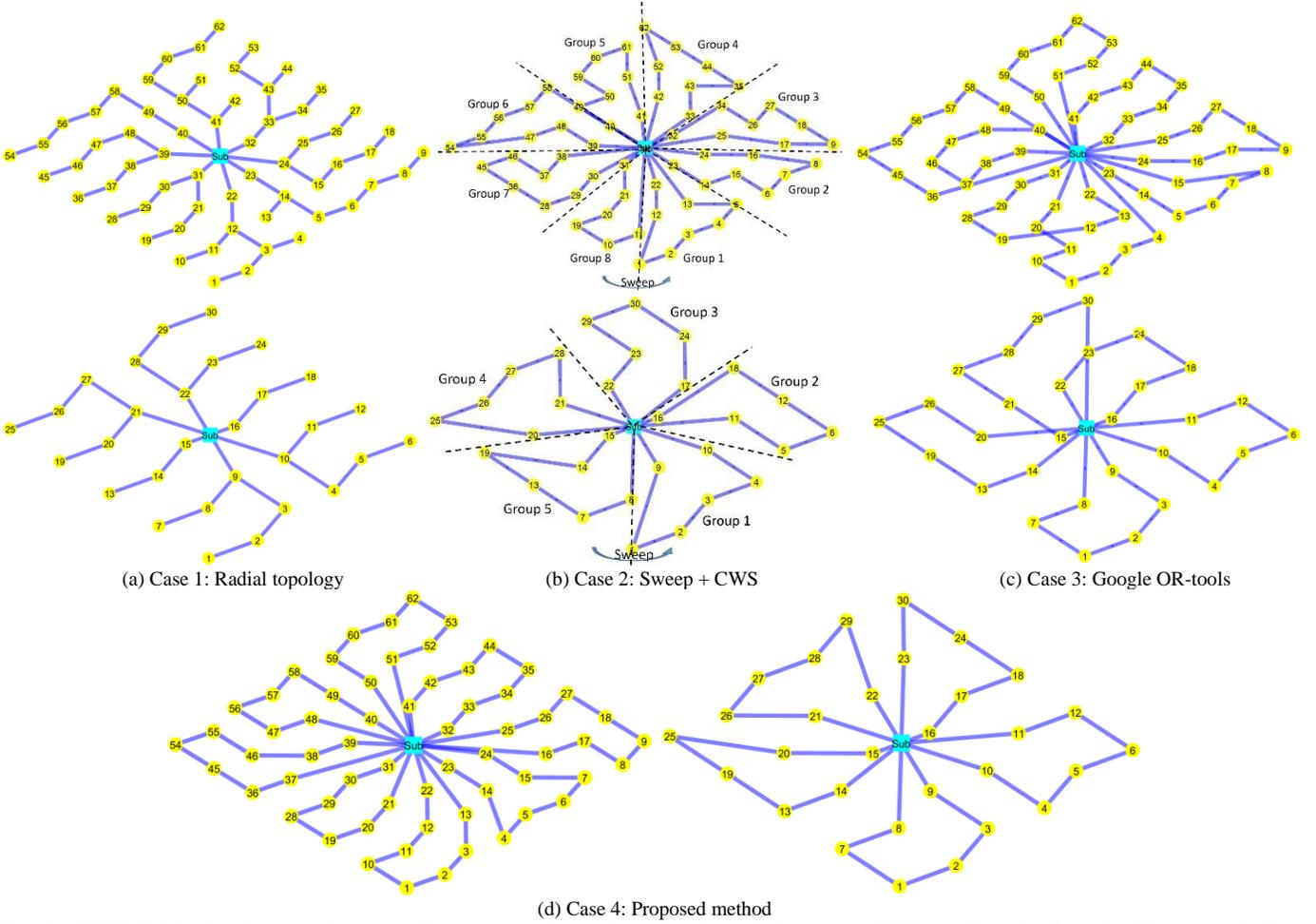

**Fig. 8.** OWF-ECS planning results in (a) Case 1: radial topology (b) Case 2: Sweep + CWS (c) Case 3: Google OR-tools (d) Case 4: proposed method for 62-WTs/30-WTs example

The OWF-ECSP results of different cases are shown in Fig. 8 and Table II. The results show that:

1) The planning results obtained in Cases 2, 3 and 4 are different even though they both use the double-sided ring topology. With the "sweep" process as shown in Fig. 8 (b) Case 2, the WTs are firstly divided into different groups and connected by the CWS in each group. Thus, connections between different groups are not allowed in Case 2. However, with Google OR-tools in Case 3 and with the proposed method used in Case 4, it's possible to find better solutions with WTs' connections in different groups.

2) After considering the cost of EENG in the lifetime of the OWF, it is beneficial to choose the double-sided ring topology (in Cases 2, 3 and 4). The investment cost of submarine cables is lowest with the radial design of the OWF-ECS (in Case 1), which is straightforward. However, the total cost of 62-WT OWF ECS (¥ 560.0 M) is much higher than those of Cases 2, 3 and 4 (¥ 445.8 M, ¥ 429.4 M and ¥ 414.8 M), reduced by 25.9% in Case 4, due to the high potential Cost of the EENG (¥ 224.1 M). However, the benefits of double-sided ring topology are not very obvious for small OWFs with less wind turbines. In the 30-WTs example, the total costs of all 4 cases are very close (¥ 193.3, 197.7, 188.7 and 190.4 M). It shows that the "N-

1" criterion might not be necessary for some small-scale OWF ECS since it could lead to some overinvestment on cables.

3) Compared with the heuristic method in Case 2, the proposed method in Case 4 saves the investment cost by improving the optimality of the OWF-ECSP solution further. In the 30-WTs example, the investment cost is saved by 3.69% (187.2→180.3). In the 62-WTs example, it's saved by 7.19% (415.9 → 386.0). Therefore, the proposed method shows significant value by finding better planning schemes for the OWF-ECS.

4) With Google OR-tools, the planning result includes some crossing cables, which are not allowed in the OWF ECS. For instance, in Fig. 8 (c), one can find crossing cables pair (19-12, 20-11) in the 62-WTs example and (21-15, 20-Sub) in the 30-WTs example. Because the Google OR-tools is based on the CVRP modeling, in which the routes of different vehicles are allowed to cross each other. In contrast, the proposed method guarantees there are no crossing cables in planning results.

5) The operation cost of the OWF-ECS, which is the cost of power losses, is significantly reduced by the double-sided ring topology design in Cases 2, 3 and 4. In the 30-WTs example, the operation cost is saved by 39.5% (16.7 in Case 1→10.1 in Case 4). In the 62-WTs example, it's saved by 44.2% (51.6 in Case 1→28.8 in Case 4).

### B. Effectiveness of cross-avoidance constraints

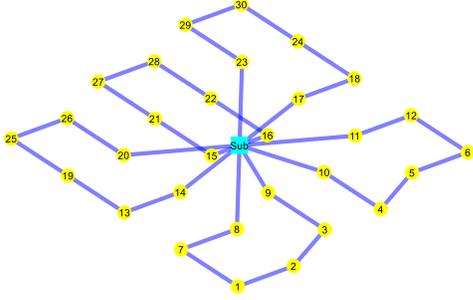

**Fig. 9**. Planning result without CAC for 30-WTs example

The cross-avoidance constraints (CAC) Eq. (8) is effective in eliminating crossing cables in the planning result. If the CAC is removed, it gets the planning result as shown in **Fig. 9**, with crossing cables pair (15-21, Sub-20), (Sub-17, 16-22) and (Sub-23, 16-22). Even though its total cost is ¥ 188.1 M, which is better than that of Case 4 in Table II (¥ 190.4 M), it's still strongly discouraged because the crossing cables are more expensive and will increase the risk of cable damages[8][13]. Similar results also happen in the 62-WTs example, but omitted here.

### C. Discussion on Computation Efficiency

We compare the proposed method with/without k-DCT model in terms of computation efficiency, the results are shown below:

Table III
OWF-ECSP Computation Efficiency with/without k-DCT

| Num of WTs | 30 | | 62 | |
|---|---|---|---|---|
| With k-DCT (Cons. (16)-(21)) | ✕ | ✓ | ✕ | ✓ |
| Total Cost (¥M) | 190.4 | 190.4 | 414.8 | 414.8 |
| Comp. Time (s) | $6.1 \times 10^4$ | 3116 | $1.6 \times 10^5$ | $8.5 \times 10^4$ |
| Optimality Gap (%) | 5.00 | 1.00 | 14.8 | 5.00 |

It's worth noting that, the proposed method does not provide a planning result proved to be global optimal even after long computation time. For instance, in the 62-WTs example, the optimality gap of the final result only reaches 5% in the end. As for the 30-WTs example, the optimality gap reaches 1.00%. Due to the dimension issue of the large-scale MIP problem (i.e., 992 continuous and 2291 binary variables for the 62-WTs example without the k-DCT model), as well as the special structure of the network design problem and the CAC, the lower bound of the relaxed problem seems to be extremely difficult to converge to a better optimality gap within acceptable time. But we believe both 1.00% and 5.00% optimality gap are acceptable results at the planning stage for OWF ECS.

Moreover, it's also worth noting that, with the complementary k-DCT model, both the 30-WTs/62-WTs test system show significant acceleration on convergence. Without the k-DCT model, it takes $6.1 \times 10^4$ s for the 30-WTs example to converge to the optimality gap 5%, and the optimality gap of the 62-WTs example is proved to be only 14.8% after $1.6 \times 10^5$

s. With the k-DCT model, after 3,116 s, the upper bound in the 30-WTs example, i.e., the current best feasible solution, is proved to be only 1.00% worse than the lower bound, and that of the 62-WTs example is proved to be 5.00% after $8.5 \times 10^4$ s, which shows good optimality gap of the planning results.

### D. Detailed Comparison with Heuristic Sweep+CWS

Since the traditional heuristic Sweep+CWS method provides different planning results when the "sweep" process starts at different wind turbines, we compare different planning results produced by the Sweep+CWS method with the proposed method below.

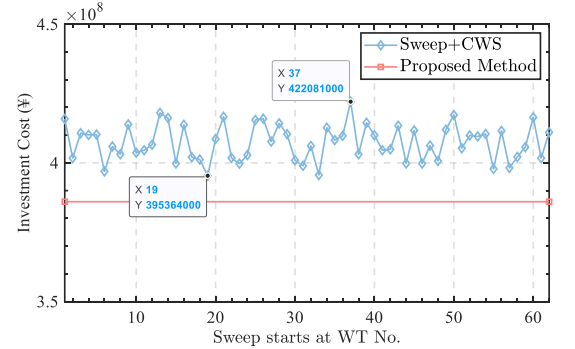

(a) 62-WTs

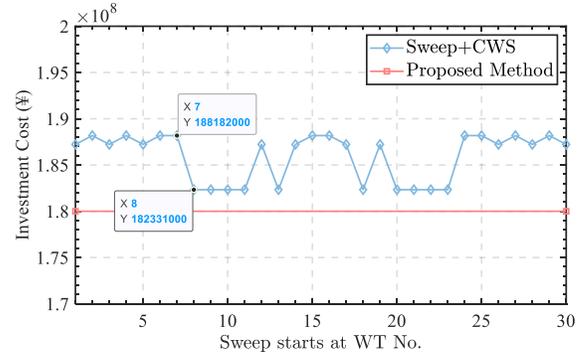

(b) 30-WTs

**Fig. 10**. Comparison on Investment Cost of (a) 62-WTs (b) 30-WTs examples with different Sweep starts WT No.

In **Fig. 10**, in the 62-WTs example, if the sweep starts at wind turbine No. 37, the investment cost of cabling is as high as ¥ 422 M, which is 9.3% higher than that of the proposed method, i.e., ¥ 386.0 M as shown in Table II. However, if the sweep starts at wind turbine No. 19, the investment cost is ¥ 395 M, which is only 2.4% higher than that of the proposed method. On average, the Sweep+CWS method provides planning results with investment cost of ¥ 407.3 M, which is 5.52% higher than that of the proposed method. As for the 30-WTs example, the investment cost ranges from ¥ 188.2 M to ¥ 182.3 M, which is 4.38%~1.11% higher than that of the proposed method. It shows that the proposed method provides more benefits for large-scale OWFs than small-scale OWFs. Moreover, the Sweep+CWS method is very efficient, the computation time for each case is only several seconds. Therefore, if the decision makers prefer computation efficiency rather than the ECS planning result's optimality, they can use the heuristic method instead of the proposed method.

## E. Error of Power Losses' Approximation

To valid the effectiveness and accuracy of using DC power flow model in OWF-ECSP, we formulate the AC power flow model in MATPOWER with all reactive power from wind turbines set to 0, and compare the error of power losses with proposed method. The results are shown in Table III.

Table IV
Error of power losses Approximation

| Num of WTs | 30 | | 62 | |
|---|---|---|---|---|
| Power losses | Case 1 | Case 2 | Case 1 | Case 2 |
| ACPF (MW) | 0.2467 | 0.1490 | 0.761 | 0.425 |
| Proposed method (MW) | 0.2458 | 0.1487 | 0.759 | 0.424 |
| Error (%) | 0.385 | 0.218 | 0.345 | 0.190 |

It could be concluded from Table IV that, with the DC power flow model used in proposed method for OWF-ECSP and the approximated power loss is included in the objective function, the error of power losses is less than 0.5%, which is ignorable in planning stage for OWF. Therefore, it's rational to apply DC power flow in proposed method considering the non-convexity of AC power flow and corresponding computation burden in optimization for OWF.

## V. Conclusion

In this paper, we propose a planning method for the offshore wind farm electric collector system (OWF-ECS) with the double-sided ring topology.

The main novelties are manifold: (i) proposing an MIQP model to optimize the cabling cost and power losses while meeting "N-1" criterion for cable failures, and (ii) developing the crossing-avoidance-constraints (CAC) to help the applicability and (iii) adding the k-degree centre tree (k-DCT) model to help accelerate the convergence and improve the global optimality of final results. The model is based on the vehicle-flow formulation used in the Capacitated Vehicle Routing Problem (CVRP) and can approximate the power losses with DC power flow in the network planning.

Case studies show the advantages of the proposed method. It can (i) significantly reduce the reliability cost compared with radial topology, (ii) eliminate planning results with crossing cables (iii) improve the solution's optimality compared with the traditional heuristic method and some software, i.e., the Sweep + Clark & Wright Saving algorithms and Google OR-tools. The disadvantage, on the other hand, is the relatively long computation time. In this sense, a hybrid optimization method combining heuristic and mathematical programming should be further explored in the future. Moreover, if the "N-1" criterion on cable failure is not necessary for the OWF, the produced planning result could be too conservative for system planner, which would lead to overinvestment on OWF ECS. Therefore, the proposed method benefits large-scale OWF more than small-scale OWF in terms of investment cost and reliability. For future work, besides the computation efficiency issue, cable sizing/OWF with multiple offshore substations should be considered, as well as planning method with detailed OWF reliability indices in the optimization model and for small-scale OWFs. Co-optimization for micro-siting and ECS planning in OWFs could also be included.

## Acknowledgment

Xinwei Shen thanks Mr. Hongke Li from PowerChina Huadong Engineering Corp. Limited (HEDC) for professional suggestions on considerations of OWF ECS planning in this paper.

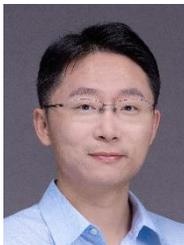

**Xinwei Shen** (M'12-SM'21) graduated from Dept. of Electrical Engineering, Tsinghua University, Beijing, China in 2010 and 2016, respectively, with B. Eng. and Ph. D. degrees. He was a visiting scholar at IIT (U. S.), U. C. Berkeley and Univ. of Macau in 2014, 2017 and 2021. He is now an Assistant Professor in Tsinghua Shenzhen International Graduate School, Tsinghua University. In 2020, he was awarded the Young Elite Scientists Sponsorship Program by Chinese Society for Electrical Engineering (CSEE). His research interests include energy internet / integrated energy system / power distribution system /ocean renewables optimization. He is the co-chair of IEEE PES Working Group on Integrated Energy System/Multi-Energy Network Modeling and Planning and officer of PES Energy Internet Coordinating Committee (EICC). He is a subject editor of *CSEE Journal of Power and Energy Systems* and young editorial board member of *Applied Energy*.

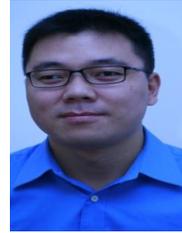

**Qiuwei Wu** (M'08-SM'15) obtained the PhD degree in Power System Engineering from Nanyang Technological University, Singapore, in 2009. He has been a tenured Associate Professor with Tsinghua-Berkeley Shenzhen Institute, Tsinghua Shenzhen International Graduate School, Tsinghua University since Jan. 2022. He was a senior R&D engineer with Vestas Technology R&D Singapore Pte Ltd from Mar. 2008 to Oct. 2009. He was working at Department of Electrical Engineering, Technical University of Denmark (DTU) from Nov. 2009 to Feb. 2022 (PostDoc Nov. 2009-Oct. 2010, Assistant Professor Nov. 2010-Aug. 2013, Associate Professor Sept. 2013-Aug. 2021). His research interests are decentralized/distributed optimal operation and control of power and energy systems with high penetration of renewables, including distributed wind power modelling and control, decentralized/distributed congestion management, voltage control and load restoration of active distribution networks, and decentralized/distributed optimal operation of integrated energy systems.

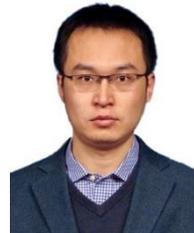

**Hongcai Zhang** (S'14-M'18) received the B.S. and Ph.D. degree in electrical engineering from Tsinghua University, Beijing, China, in 2013 and 2018, respectively. He is currently an Assistant Professor with the State Key Laboratory of Internet of Things for Smart City and Department of Electrical and Computer Engineering, University of Macau, Macao, China. In 2018-2019, he was a postdoctoral scholar with the Energy, Controls, and Applications Lab at University of California, Berkeley, where he also worked as a visiting student researcher in 2016. His current research interests include integrated energy systems, transportation electrification, and distributed energy resources.

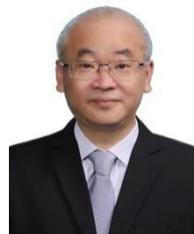

**Liming Wang** (M'10-SM'18) was born in Shaoxing, Zhejiang Province, China, on 30 November 1963, and received the B.S., M.S., and Ph.D. degrees in high voltage engineering from the Department of Electrical Engineering, Tsinghua University, Beijing, P.R. China, in 1987, 1990, and 1993, respectively. He has worked at Tsinghua University since 1993. His major research fields are high voltage insulation and electrical discharge, flashover mechanism on contaminated insulators, and application of pulsed electric fields.